\documentstyle[preprint,aps]{revtex}
\newcommand{\be}{\begin{equation}}
\newcommand{\la}{\label}
\newcommand{\ee}{\end{equation}}
\newcommand{\bea}{\begin{eqnarray}}
\newcommand{\eea}{\end{eqnarray}}
\newcommand{\rp}{\dot{R}}
\newcommand{\gp}{G^\prime}

\begin{document}

\title{Analytical solution of the dynamical spherical MIT bag}
\author{K.~Colanero, M.~-C.~Chu}
\address{Department of Physics, The Chinese University of Hong Kong, 
Shatin, N.T., Hong Kong.}
\maketitle
\begin{abstract}
We prove that when the bag surface is allowed to move radially,
the equations of motion derived from the MIT bag Lagrangian 
with massless quarks and a spherical boundary admit only one 
solution, which corresponds to a bag expanding at the speed of 
light.  This result implies that some new physics ingredients, such as coupling
to meson fields,  are needed to
make the dynamical bag a consistent model of hadrons.   
\end{abstract}

\pacs{PACS number(s): 12.39.Ba, 03.65.Pm, 03.50.Kk}  

\narrowtext

The MIT bag model, in which hadrons are modelled by the states formed with
free quarks confined inside an impenetrable bag, has been rather 
successful in reproducing static ground state properties of 
hadrons \cite{MIT}.  
Because of its simplicity, especially its spherically symmetric version, 
the model has been used extensively in the discussion of various 
phenomena ranging from strange stars \cite{sstar} to ultra-relativistic 
heavy-ion 
collisions \cite{rhic}, even though these often involve situations of 
high density/temperature where the applicability of the model is doubtful.
Many attempts have also been made to augment the basic MIT bag model
with new ingredients, such as the partial restoration of chiral symmetry 
via
meson coupling \cite{chbag} and inclusion of perturbative gluon exchanges 
among quarks \cite{gluonex}.  However, in almost all discussion, the 
hadron bag
is treated as a static boundary between the perturbative and 
nonperturbative vacuum, and excitations of the hadron are associated with 
the quark degree of freedom.  
The few notable exceptions \cite{rebbi,kuti,nogami,brown,meissner},
which allowed for the possibility of
a dynamical bag boundary, focused mainly on reproducing the correct 
phenomenological parity order of the low-lying states of the nucleon,  
but several approximations and modifications to the theory had to 
be employed.
Moreover the more fundamental
question of whether it is consistent and feasible to use a dynamical bag
to model hadrons was not addressed.  That is the motivation of 
the present work. 
In this paper we discuss, analytically and without approximations,  
the consequences of allowing the bag boundary to move.  

We consider the basic MIT bag model with free fermions inside 
a spherically symmetric but non-static bag.  We will show that the 
equations of motion require that the 
fermion field $\psi(t,r)$ vanishes at the bag boundary $r=R(t)$. 
Hence our problem reduces to a quite general one: that of the 
Dirac equation in a spherical dynamical cavity with $\psi(t,R(t))=0$.
Furthermore, we will show that for massless fermions, the {\it only}
solution is that of a bag expanding with the speed of light. 
From this unexpected result, which evidently has no phenomenological
application,
we infer that some new physics ingredients, such as an interaction
term with other fields,
have to be introduced to make the more general dynamical bag model
consistent and physical \cite{klaus}.
In the case of massive quarks we could not find a solution and we 
conjecture that in fact it does not exist.

The MIT bag Lagrangian density \cite{MIT} is written as
\be
{\cal L}=\left[{i\over 2}\left(\bar\psi \gamma^\mu \partial_\mu \psi - 
\partial_\mu \bar\psi \gamma^\mu \psi\right)
- B\right] \theta_v (x) - {1\over2}\bar\psi \psi \Delta_s  \; ,
\la{mitlag}
\ee
where $\theta_v (x)$ is $1$ inside the bag and $0$ outside and 

\be
{\partial \theta_v \over \partial x^\mu} = n_\mu \Delta_s \; ,
\la{dthetadmu}
\ee
$\Delta_s$ being the surface delta-function and $n_\mu$ the normal vector 
to the bag. From the Euler-Lagrange equation of motion

\be
{\partial{\cal L} \over \partial \bar\psi}-\partial_\mu{\partial {\cal L}
\over \partial \left(\partial_\mu \bar\psi\right)} = 0 \ \ ,
\la{eulerlag}
\ee
we obtain

\be
i \gamma^\mu \partial_\mu \psi = 0 \ , \hspace{1 cm} \mbox{inside 
the bag} \; ,
\la{eqinside}
\ee
\be
i\gamma^\mu n_\mu \psi = \psi \ , \hspace{1 cm} \mbox{on the surface} \; .
\la{linearbc}
\ee
This last equation may be considered as the boundary condition for 
Eq.~\ref{eqinside}. Energy-momentum conservation  
implies a further constraint at the boundary \cite{MIT,bhaduri}:

\be
B n^\nu = {1\over 2} [\partial^\nu \left(\bar\psi \psi\right) ]_{r=R} \; .
\la{nonlinearbc}
\ee

We now look for a spherically symmetric solution of the above equations. 
In this case we have
$\theta_v = \theta(R-r)$, $\Delta_s = \delta(R-r)$, and
\be
n_\mu = (\rp , -\hat{r}) \; .
\la{nmuss}
\ee
We first find an explicit expression for the boundary condition. 
Eq.~\ref{linearbc} becomes 
\be
i \rp \gamma^0 \psi -i \vec{\gamma}\cdot \hat{r} \psi = \psi \ \ ,
\la{linearbc1}
\ee
with
\be
\vec{\gamma}\cdot \hat{r} = \left ( \begin{array}{cc}
0 & \vec{\sigma} \cdot \hat{r} \\
- \vec{\sigma} \cdot \hat{r} & 0
\end{array} \right ) \; .
\la{gammar}
\ee
We can write the spinor $\psi$ as \cite{bhaduri,bogo}
\be
\psi = \left ( \begin{array}{c}
\phi \\
\chi
\end{array} \right ) = \left ( \begin{array}{c}
g(r,t) {\cal Y}^{j_3}_{jl} \\
i f(r,t) {\cal Y}^{j_3}_{jl^\prime}
\end{array} \right ) \ \ ,
\la{psiofr}
\ee
where ${\cal Y}^{j_3}_{jl}$ contains the spin and angular parts of the 
wave function.  Observing that $(\vec{\sigma} \cdot \hat{r}) 
{\cal Y}^{j_3}_{jl} = - {\cal Y}^{j_3}_{jl^\prime}$ \cite{bhaduri},
we can write Eq.~\ref{linearbc1} as 
\be
\left \{ \begin{array}{lll}
i\rp g(t,R) - f(t,R) & = & g(t,R) \ ,\\
\rp f(t,R) - i g(t,R) & = & i f(t,R) \ .
\end{array} \right .
\la{linearbc2}
\ee
If $\rp=0$ we have the familiar boundary condition for the static MIT 
bag, 
{\it i.e.}~$g(t,R)=-f(t,R)$, which
also corresponds to the one in the Bogolioubov model \cite{bogo}, 
and the analytical solutions are well known. 
However it is easy to verify that, if $\rp \neq 0$,
then Eqs.~\ref{linearbc2} can be satisfied only if 
\be
g(t,R)=f(t,R)=0 \; . 
\la{linearbc3}
\ee
Notice from Eq.~\ref{nonlinearbc} that this implies $B=0$, 
and in this way energy-momentum is conserved regardless 
of the motion of the bag.
This also means that Eq.~\ref{nonlinearbc} does not provide information 
about $\rp$.

Performing the change of variable $y=r R_0/R(t)$ one could recast the 
problem into a static boundary one.
In this framework the motion of the bag's surface is treated 
as a time dependent perturbation to the static Hamiltonian and one 
looks for the solutions by means of a time dependent expansion 
in terms of the static 
cavity eigenfunctions, $\exp\{-iE_nt\}\psi_n (y)$ \cite{schrod}.
However in our case this approach cannot provide us the solution. In fact,
writing the wave function $\psi_f$ for the fixed boundary problem as
\be
\psi_f=\sum_{n=0}^{\infty} c_n(t) e^{-iE_n t} \psi_n (y) \; ,
\la{psiexp}
\ee
we can in general work out $c_n(t)$, {\it e.g.}~by perturbation theory, 
but it is well known that the static eigenfunctions, 
$\exp\{-iE_nt\}\psi_n (y)$, of the Dirac field inside
a spherical cavity are non-zero at the boundary \cite{bhaduri}, and 
we have no way to impose the boundary condition Eq.~\ref{linearbc3} 
on expression Eq.~\ref{psiexp}.
In other words, although at the initial time we can always choose a 
suitable combination of $\psi_n (y)$ which is zero at $y=R_0$, 
at subsequent times $\psi_f(t,R_0)$, as expressed in Eq.~\ref{psiexp},
will in general be different from zero.
We hence need to proceed in a different way.

Substituting expression Eq.~\ref{psiofr} for $\psi$ in Eq.~\ref{eqinside},
with $l=0$ and $l^\prime = 1$ in order to have spherical symmetry,
we obtain two coupled equations:

\be
-i {\partial g \over \partial t} = {\partial f \over \partial r} + 
{2\over r} f \ \ ,
\la{eqinside1}
\ee
\be
i {\partial f \over \partial t} = {\partial g \over \partial r} \; .
\la{eqinside2}
\ee
Integrating Eq.~\ref{eqinside2} with respect to time and substituting 
the expression for $f(t,r)$ in Eq.~\ref{eqinside1} we have

\be
f = A(r)-i \int dt {\partial g \over \partial r}  
\la{eqinside4}
\ee
\be
-i {\partial g \over \partial t} = -i \int dt \left[{\partial^2 g \over 
\partial r^2} + {2\over r} {\partial g \over \partial r}\right]
+{d A(r) \over dr} + {2\over r} A(r) \; ,
\la{eqinside3}
\ee
where $A(r)$ is a time independent function to be determined. 
Eq.~\ref{eqinside3} is the spherical wave equation in 
integro-differential form, whose general solution can be written 
in the form:
\be
g(t,r)={1 \over r}\left[G(t-r) - G(t+r) \right] + C t + D \ \ ,
\la{g(t,r)}
\ee
with $C$ and $D$ two constants.
In principle, a term of the form $\alpha/r$ is allowed in Eq. 18.  
However, it makes the wavefunction
not normalizable and unphysical. We have
also checked that this term does not affect at all our proof. We 
therefore set $\alpha = 0$.

Inserting the above expression in 
Eq.~\ref{eqinside3} yields an equation for $A(r)$:
\be
{d A\over dr}=-{2\over r} A -i C \ \ ,
\la{eqab}
\ee
whose solution is
\be
A(r)=-{i\over 3}C r \; .
\la{A(r)}
\ee
Hence we can write an explicit expression for $f$ as
\be
f(t,r) = {i\over r} \huge\{ G(t-r)+G(t+r)+
{1\over r}\left[ Q(t-r)-Q(t+r) \right] \huge\} - i {C\over 3} r  \ \ , 
\la{f(t,r)}
\ee
with $Q^\prime (z) = G(z)$.
Now using the boundary conditions Eq.~\ref{linearbc3} we can derive the 
relations among $G(t-R)$, 
$G(t+R)$, $R(t)$ and $\rp(t)$. Equating $g(t,R)$ and $f(t,R)$ to zero we 
have
\be
g(t,R)={1 \over R}\left[G(t-R) - G(t+R) \right] + C t + D = 0 \; ,
\la{g(t,R)=0}
\ee
\be
f(t,R) = {i\over R} \huge\{ G(t-R)+G(t+R)+
{1\over R}[Q(t-R)-Q(t+R)] \huge\} - i {C\over 3} R = 0 \; .
\la{f(t,R)=0}
\ee
Taking the time derivative of both equations we obtain
\be
{d \over dt}g(t,R) = {1\over R} \left\{ (1- \rp) \gp(t-R) - 
(1+\rp) \gp(t+R) \right\}
- {\rp\over R^2}\left[G(t-R) - G(t+R) \right] + C = 0
\la{dgdt=0}
\ee
and
\be
\begin{array}{lll}
& &{d \over dt}f(t,R) = -{\rp\over R}f(t,R) - {i\over R} \Huge\{ 
{2\over 3}C R \rp -
(1-\rp ) \gp (t-R) - (1+ \rp ) \gp (t+R) +  \\
& & \\
& &{\rp \over R^2} \left[Q(t-R)-Q(t+R)\right] +
{1 \over R}\left[\left( \rp-1 \right)G(t-R) + \left(\rp+1 \right)G(t+R)
\right] \Huge\} 
= 0 \; .
\end{array}
\la{dfdt=0}
\ee
From the last two equations, and by means of Eqs.~\ref{g(t,R)=0}~
and~\ref{f(t,R)=0}, we can find
\be
(1-\rp ) \gp (t-R) - (1+\rp ) \gp (t+R)
+\rp (Ct+D) + C R = 0
\la{eq1}
\ee
and
\be
- (1-\rp ) \gp (t-R) - (1+\rp ) \gp (t+R)
+ Ct + D + C R \rp = 0 \; .
\la{eq2}
\ee
These two equations contain all the information we need to solve the 
problem, 
{\it i.e.}, to find $\rp (t)$ for any 
$t > t_0$ and $G(z)$ for any $z > t_0 + R(t_0)$. 
In fact, using the same argument as in \cite{colchu}, as long as 
$| \rp | \leq 1$, at each time $t\geq t_0$ $\gp (t-R)$
is known and we have two equations for the two unknowns $G(t+R)$ and $\rp 
(t)$.
By summing and subtracting Eqs.~\ref{eq1}~and~\ref{eq2} we can decouple 
$\rp$ and $\gp (t+R)$ as follows
\be
\left(1+\rp \right) \left[ -2 \gp (t+R) + (Ct+D+CR) \right] =  0 \; ,
\la{eq3}
\ee
\be
\left(1-\rp \right) \left[ 2 \gp (t-R) - (Ct+D-CR) \right] =  0 \; .
\la{eq4}
\ee
It is finally evident that Eq.~\ref{eq4} implies
\be
\rp(t)=1 \ \ ,
\la{rp=1}
\ee
and from Eq.~\ref{eq3} we have
\be
\gp \left( t+R(t) \right) = {1 \over 2}\left[Ct+D+CR(t) \right] \; .
\la{gp(t+R)}
\ee

Notice that $\rp = -1$ is not allowed because Eq.~\ref{eq4} would not
be satisfied. Since expressions Eqs.~\ref{g(t,r)}~and~\ref{f(t,r)} 
represent the general solution of the problem, the above result 
excludes the possibility of any other solution. It is also important 
to note that, with $\rp =1$, $t-R(t)=t_0-R(t_0)$.

Now, knowing that $R(t) = R_0+t-t_0$ and defining $z \equiv t+R(t)$, 
we have
\be
\gp (z)={1\over 2} (Cz+D) \hspace{1 cm} z\geq t_0+R_0 \; .
\la{gp(z)}
\ee
At this point, in order to have a clear understanding of the solution, 
we set $C=D=0$.  In fact, the two constants, with the boundary conditions 
given by Eqs.~\ref{g(t,R)=0}~and~\ref{f(t,R)=0}, are physically
irrelevant. We obtain
\be
G(z)=G(z_0)=G(t_0-R_0)  \hspace{1 cm} z\geq t_0+R_0=z_0 \; ,
\la{G(z)=const.}
\ee
\be
Q(z)= G(z_0) (z-z_0) + Q(z_0) \hspace{1 cm} z\geq t_0+R_0=z_0 \; .
\la{Q(z)=}
\ee
Given $\psi(t_0,r)$, which for example is different from zero at some 
$r$, 
then $\psi(t,r)$ will go to zero as $t \rightarrow t_0+R_0+r$, and 
so we will have
\be
\psi(t,r)=0 \ , \hspace{1 cm} r\leq t-t_0-R_0 \; , \;\; r\geq t-t_0+R_0 
\; .
\la{psi(t,r)}
\ee

Therefore, after a time $t=R_0+t_0$, the solution represents an expanding 
spherical shell with internal radius $R_{\rm in}=t-t_0-R_0$ and 
external radius $R_{\rm out}=t-t_0+R_0$.
The solution for $t<t_0$ can be found analogously, evolving backward in 
time. In this case Eqs.~\ref{eq3}~and~~\ref{eq4}
imply $\rp=-1$ and $\gp(t-R)$ would be the unknown. 
Obviously the solution becomes singular as $R(t)=0$.

This unexpected solution of the massless Dirac equation in a spherical 
dynamical cavity is evidently due to the boundary conditions 
Eq.~\ref{linearbc3}. However, as far as one considers the Dirac field 
completely confined in a cavity, any other boundary condition would 
violate unitarity, as can be checked easily by taking the time derivative 
of the norm of the field.

From a physical point of view the problem lies in the fact 
that the MIT bag model sets the field to zero outside the bag already in 
the 
Lagrangian. In the Bogolioubov model with a finite
square well potential, there is a non-zero field also outside the well. 
If the wall moves inward, the field
gains enough energy so that parts of it can go out of the well. 
The higher the potential the more energy is 
transferred to the field during a compression, mainly because the
Dirac field at the wall does not approach zero as the potential goes to 
infinity.

The MIT bag model overcomes this problem by means of the 
boundary condition Eq.~\ref{linearbc3}, which sets the field to
zero at the moving boundary. The drawback, as we have seen, is the 
absence of a bag-like solution for a dynamical boundary.

We also asked ourselves whether giving a mass to the fermions would 
alleviate the problem. The static bag wavefunctions, as their massless 
counterparts, do not equal zero at the boundary.
A moving boundary, however, requires also in the massive case 
the boundary conditions Eq.~\ref{linearbc3} and hence, as mentioned 
before,
the solution cannot be written as a time dependent combination 
of static cavity eigenfunctions. We could not find the general solution 
in this case, and we conjecture that no solution exists for massive 
fermions.
Our conjecture is based on the fact that Eq.~\ref{linearbc3} implies
a delicate cancellation of all the Fourier components of the wavefunction.
In order to maintain this zero boundary condition at all times, 
all the Fourier components of the wave should travel at the
same speed as that of the moving wall.  This is indeed possible in 
the massless case, if the bag wall expands with the speed of light.
For massive fermions however, each Fourier component travels at a 
different speed, and it may not be possible to satisfy the boundary
condition Eq.~\ref{linearbc3} at all times.

In this paper we have considered specifically the MIT bag
model, which is singular in the sense that no kinetic energy
term is associated with the motion of the boundary. However it is
straightforward to see that our result applies without modification
to a non-singular model such as the ``Budapest'' bag \cite{kuti}
which includes a surface tension besides the volume energy term.
This is because the presence of a boundary kinetic  
term does not modify the linear boundary condition Eq.~\ref{linearbc},
which is responsible for the unphysical solution.
Analogously to the MIT bag, energy conservation would require the
``Budapest'' bag to have a zero surface tension, besides $B=0$.

In summary,
we have shown that 
the equations of motion derived from the {\it dynamical} MIT bag Lagrangian 
with massless quarks and a spherical boundary admit only one 
solution corresponding to a bag expanding at the speed of 
light.  This result raises the question of whether the quantization of 
the theory can provide stable solutions.  We infer that
a consistent dynamical bag model for 
absolutely confined fermions must include an interaction term
with some other fields at least at the boundary of the domain.
For example, in a work to be published elsewhere \cite{klaus} we will 
show that a {\it dynamical}
chiral bag model \cite{chbag} admits physically meaningful
solutions.

We thank the support of a Hong Kong Research Grants Council grant CUHK
4189/97P and a Chinese University Direct Grant (Grant No.~2060193).


\end{document}